# Design and proof of concept for silicon-based quantum dot quantum bits

[1,2]Mark Friesen, [2]Paul Rugheimer, [2]Donald E. Savage, [1,2]Max G. Lagally, [3]Daniel W. van der Weide, [1]Robert Joynt, and [1]Mark A. Eriksson

[1]*Department of Physics, University of Wisconsin, Madison, Wisconsin 53706, USA*
[2]*Department of Materials Science and Engineering, University of Wisconsin, Madison, Wisconsin 53706, USA*
[3]*Department of Electrical and Computer Engineering, University of Wisconsin, Madison, Wisconsin 53706, USA*

Correspondence and requests for materials should be addressed to M. F. (email: friesen@cae.wisc.edu).

**Abstract:** Spins based in silicon provide one of the most promising architectures for quantum computing. Quantum dots are an inherently scalable technology. Here, we combine these two concepts into a workable design for a silicon-germanium quantum bit. The novel structure incorporates vertical and lateral tunneling, provides controlled coupling between dots, and enables single electron occupation of each dot. Precise modeling of the design elucidates its potential for scalable quantum computing. For the first time it is possible to translate the requirements of fault-tolerant error correction into specific requirements for gate voltage control electronics in quantum dots. We demonstrate that these requirements are met by existing pulse generators in the kHz-MHz range, but GHz operation is not yet achievable. Our calculations further pinpoint device features that enhance operation speed and robustness against leakage errors. We find that the component technologies for silicon quantum dot quantum computers are already in hand.

Quantum computing offers the prospect of breaking out of the classical von Neumann paradigm that dominates present-day computation. It would enable huge speedups of certain very hard problems, notably factorization[1]. Constructing a quantum computer (QC) presents many challenges, however. Chief among these is scalability: the $10^6$ qubits needed for simple applications[2] far exceed the potential of existing implementations. This requirement points strongly in the direction of Si-based electronics for QC. Silicon devices offer the advantage of long spin coherence times, fast operation, and a proven record of scalable integration.

Specific Si-based qubit proposals utilize donor-bound nuclear[3,4] or electronic[5] spins as qubits. However, quantum dots can also be used to house electron spins[6-8], and they have the advantage that the electrostatic gates controlling qubit operations are naturally aligned to each qubit. These proposals describe an intriguing possibility. Our aim here is to describe a new SiGe qubit design, and, just as importantly, to carry out detailed modeling of a specific design for the first time. Modeling provides a proof of principle, pinpoints problem areas, and suggests new directions.

The fundamental goal of our design is the ability to reduce the electron occupation of an individual dot precisely to one, as in vertically coupled structures. It may be possible to use the spin of multi-electron quantum dots as qubits, but single occupation is clearly desirable. The spin state "up" = $|0\rangle$ or "down" = $|1\rangle$, stores the quantum bit of information. At the same time, it is necessary to have tunable coupling between neighboring dots. This is achieved by controlled movement of electrons along the quantum well that contains two dots. The solution is to draw on two distinct quantum dot technologies: lateral and vertical tunneling quantum dots[9].

The design, shown in Fig. 1, incorporates a *back-gate* that serves as an electron reservoir, a *quantum well* that confines electrons vertically, and split *top gates* that provide lateral confinement by electrostatic repulsion. All semiconductor layers are formed of strain-relaxed $Si_{1-x}Ge_x$ except the quantum well, which is pure, strained Si. Relaxation is achieved by step-graded compositional growth on a Si wafer[10]. Here, we consider the composition $x = 0.077$, consistent with a quantum well band offset $\Delta E_c \cong 84$ meV, with respect to the barriers.

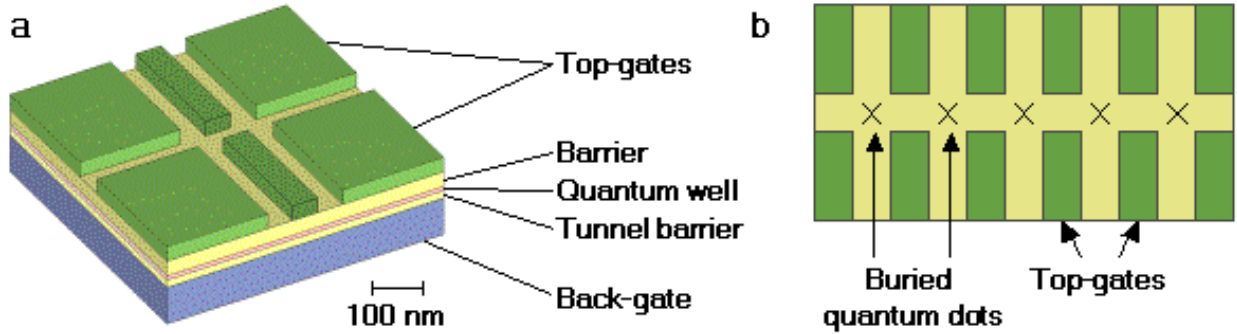

**Figure 1** Two implementations of a quantum dot quantum computer. The cross-section of the heterostructure shows, from bottom to top, a thick, *n*-doped, strain-relaxed $Si_{1-x}Ge_x$ back-gate (blue), a 10 nm undoped $Si_{1-x}Ge_x$ tunnel barrier (yellow), a 6 nm undoped Si quantum well (red), a 20 nm undoped $Si_{1-x}Ge_x$ barrier (yellow), and lithographically-patterned metallic top-gates (green). All fabrication steps are based on standard technology. Not pictured is a thin Si capping layer. **a,** shows a side-view of the double quantum dot structure studied in this work. **b,** shows a top-view of a multidot device, demonstrating the scalability.

The novel feature of our design is the combined use of vertical tunnel coupling through the back gate, together with lateral coupling defined by the split top gates. To load a single electron into a dot, the gate potentials are adjusted so that single-electron filling is energetically favored. The energetic stability of such single-electron filling is computed below. In contrast with conventional, laterally coupled dots, we emphasize that no electrical connections are made to the quantum well layer. We shall limit our discussion in this paper to two qubits, though clearly the design is scalable. Two qubits are enough for implementation of all necessary gates. Readout requires more qubits and separate consideration[3-8].

The time evolution of the qubits is controlled by the exchange interaction, which is the spin-dependent part of the Coulomb interaction. For a two-electron system the interaction can be expressed as $H_s(t) = J(t)\mathbf{S}_1 \cdot \mathbf{S}_2$, with the time-evolution operator, $U(t) = \exp[i\int H_s(\mathbf{t})d\mathbf{t}/\hbar]$. The exchange coupling, *J*, is only appreciable when the electron wavefunctions overlap. It can be extinguished by raising an electrostatic barrier between the electrons, forcing them to separate. By using a coded qubit scheme[11,12], exchange coupling becomes the basis for both two-qubit gate operations like SWAP, as well as one-qubit operations like rotations. Since all quantum gates, including the controlled (C)-NOT, can be expressed as combinations of such basic operations[6,7], the exchange coupling becomes a universal control element for quantum dot quantum computing[11,12].

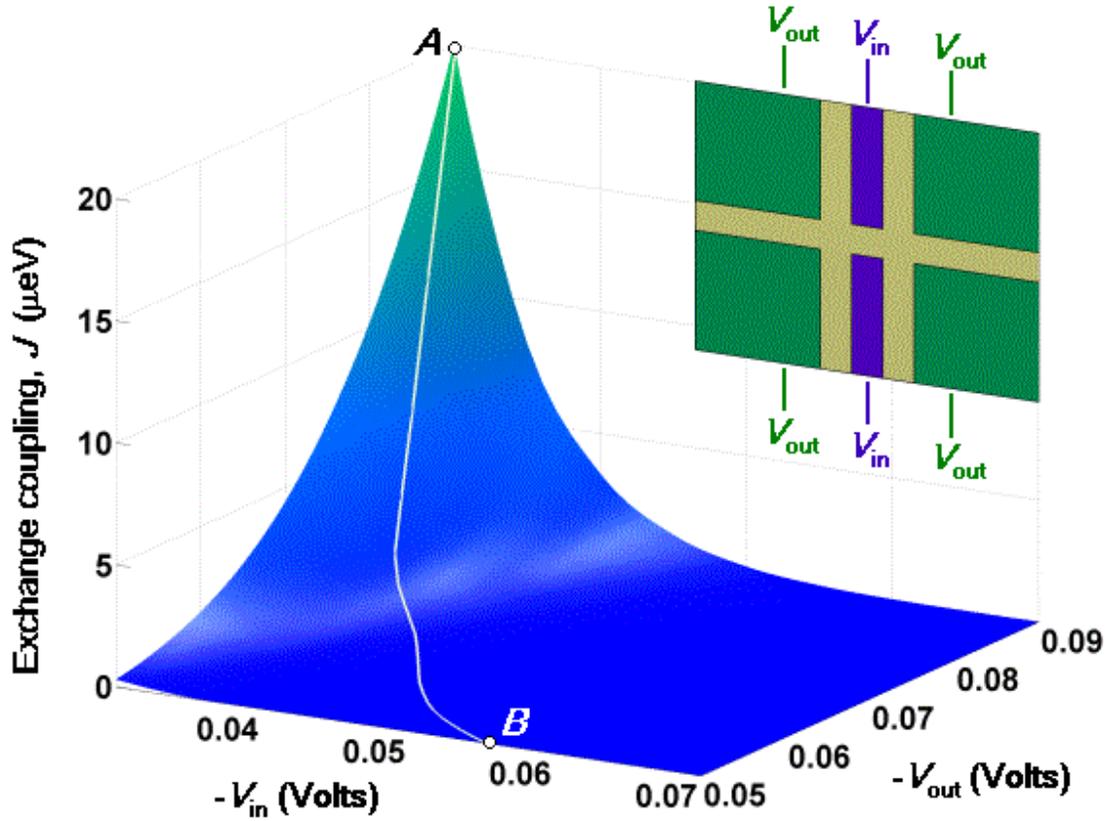

**Figure 2** Exchange coupling, $J$, computed for the double quantum dot device shown in Fig.1a. The top-gate potentials, $V_{out}$ and $V_{in}$, are explained in the inset, while the back-gate is held to ground. Curve $AB$ marks the line of maximum stability for gate operation. (See Fig. 4.)

The mapping $J(V_1, V_2, ...)$ between the top-gate voltages and the exchange coupling succinctly characterizes the operation of the quantum computer. We have computed this mapping numerically for the device shown in Fig. 1a, obtaining exact solutions for the case of one electron per dot. The exchange coupling is calculated as the difference between the ground and first excited states energies: $J = E_{trip} - E_{sing}$ [6,7]. "Singlet" and "triplet" refer to the spin symmetry of the two-electron wavefunction. Significant numerical accuracy is required in the calculations because of the large difference in energy scales: $J / E_{trip} < 5 \times 10^{-4}$.

Figure 2 shows a map of the exchange coupling $J$ as a function of gate voltages. As we show below, Fig. 2 enables us to calculate error rates in the quantum gates due to noise or uncertainty in the gate voltages. To simplify the analysis, we have considered only two independent gate voltages, $V_{out}$ and $V_{in}$, corresponding to voltages on the outer four and inner two top-gates, respectively. (See inset.) The back-gate is set to ground. The trends observed in

Fig. 2 are consistent with previous studies, which use more idealized confinement potentials[6,7,13]. For the case of zero magnetic field, studied here, the exchange coupling does not cross zero, in contrast with predictions for high fields[6,7,13]. Thus exchanged-based gating can never be switched off completely. However $J$ can always be made arbitrarily small, by creating a large barrier. Indeed, this may be the most robust technique for controlling switch-off errors.

Figure 3 provides insight into the operation of the device. Results are shown for two configurations of the gate voltages, corresponding to points $A$ and $B$ in Fig. 2. For case $A$, corresponding to a low barrier ($|V_{in}| << |V_{out}|$), the potential landscape becomes an elongated trough, with significant overlap of the electronic wavefunctions and a large exchange coupling ($J \cong 20\ \mu eV$). For case $B$, corresponding to a high barrier ($|V_{in}| >> |V_{out}|$), the potential wells and the electrons are separate and distinct, with a vanishing exchange coupling. Because of the proximity of the electrons to the back-gate ($\sim 13\ nm$), the image charges in the conducting back-gate have a stronger effect than those in the top-gates. Consequently, screening of the Coulomb interactions occurs over length scales greater than 26 nm. The net effect is to amplify the features of the potential landscape in Fig. 3, as well as the switching characteristics of $J$. Hence, image charges are actually desirable, and control over screening forms a useful design tool.

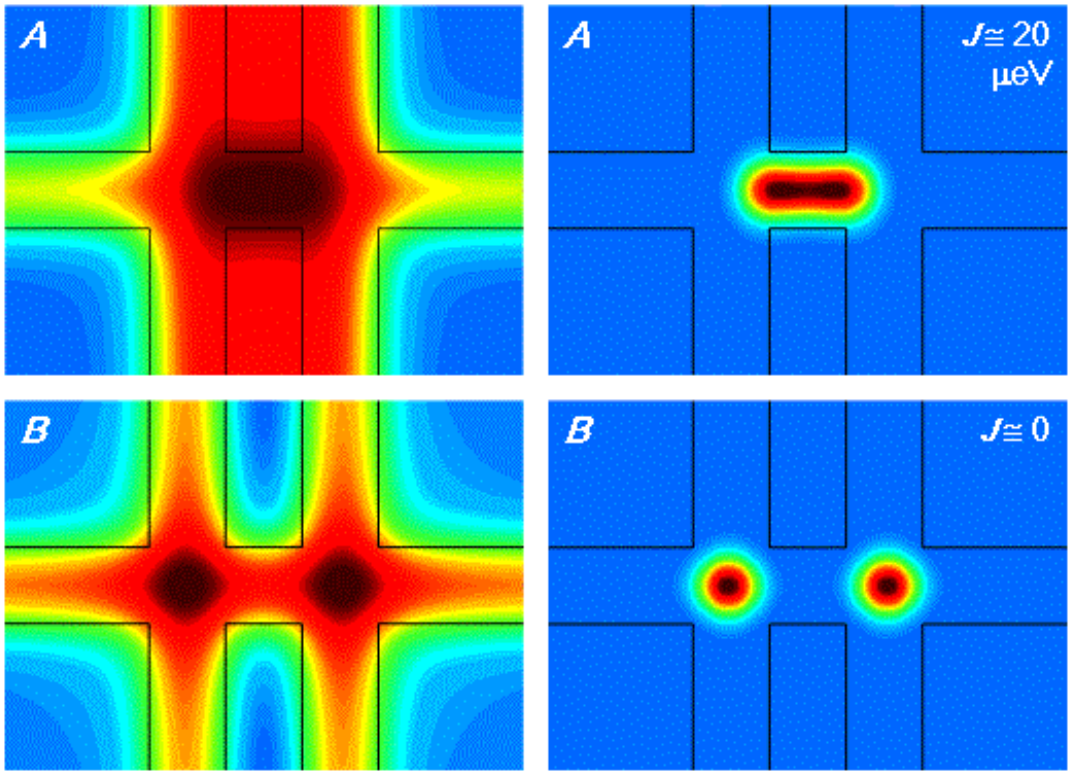

**Figure 3** Electrostatic potentials (left) and charge density maps (right), as computed in the center of the quantum well. Cases $A$ and $B$ correspond to the endpoints of the curve $AB$ shown in Fig. 2. The calculated exchange coupling, $J$, is given for both cases.

Errors will arise during the course of a quantum computation, some of which can be repaired, while others cannot. An example of the latter is "leakage," in which a trapped electron is excited into a mode outside the qubit Hilbert space. This may happen, for example, by a transition of an electron to an orbitally excited state of the dot. Since most leakage cannot be corrected by quantum error correction (quantum "software"), it must be controlled via hardware design. The idea is to create an environment where the most dangerous excitations (low-lying orbital modes) occur well above the thermal energy scale. Fortunately, most excitations are separated from the ground state by at least 1 meV (12 K). For example, the closest lying orbital states are 1.5 meV (17 K) above the ground state, and do not pose a significant leakage danger for low temperature operation. However, there is a splitting of the ground state due to a weak conduction valley-orbit coupling. This valley-orbit coupling arises from the presence of a quantum well and an applied electric field from the top and back-gates[14]. Such valley interference effects are known to cause erratic behavior in $J$ [15]. For the device studied here, the splitting is of order 0.06 meV (0.7 K), but can be increased by a factor of $5-10$ by modifying the heterostructure architecture to allow larger electric fields. Thus, at sufficiently low temperatures, all orbital leakage is suppressed.

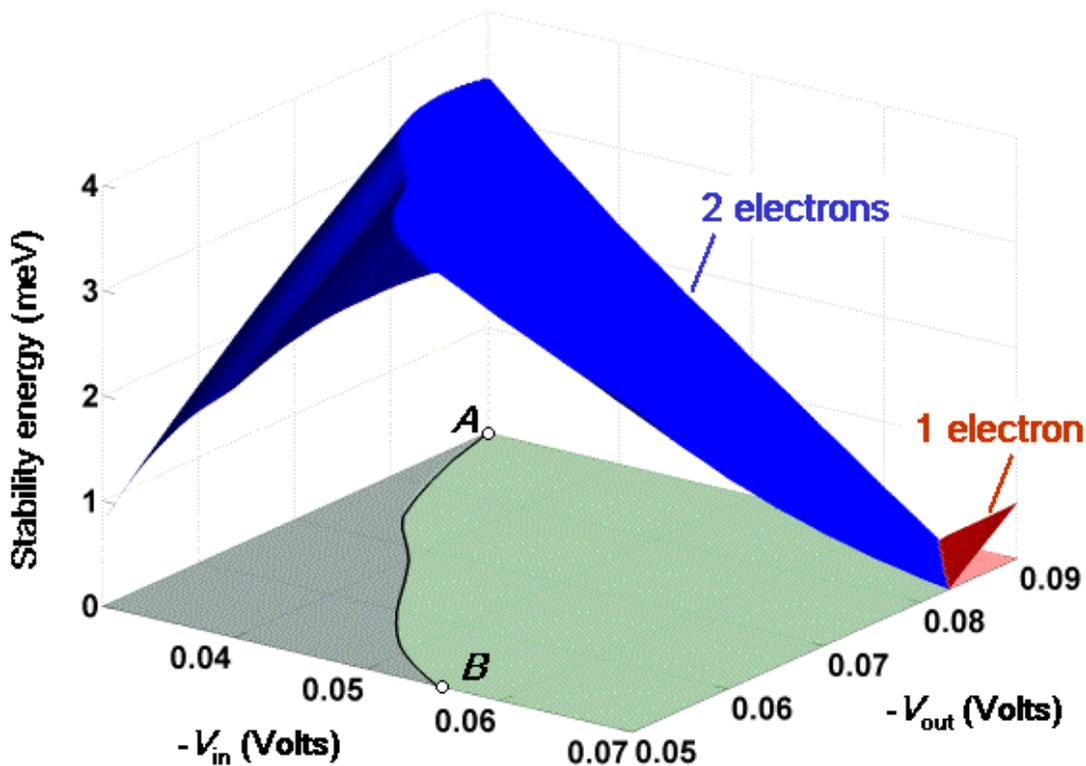

**Figure 4** Excitation energy (stability energy) for changing the electron filling number in the quantum dot, plotted against gate potentials. The two-electron stability range is shown in blue, while the one-electron stability range is shown in red. Three-electron stability does not occur in the range of voltages shown here. Optimal two-electron stability occurs along curve *AB*.

A second type of leakage error for spin qubits is associated with electron tunneling. The number of electrons per dot must remain fixed. In our device, the same gates that control the electron positions and the exchange coupling can also raise or lower the energy of the quantum well with respect to the tunnel-coupled back-gate. Normal gating procedures might therefore cause leakage. In Fig. 4 we have plotted the excitation energy for an electron in a double dot to tunnel into (or out of) the back-gate. The desired filling number for the double-dot (one electron per dot) remains stable over a wide range of gate voltages, with stability energies up to $3-4$ meV. Tunneling is therefore not a particularly dangerous form of leakage. The band offset, $\Delta E_c$, which is determined by the Ge content of the barriers, is an important, tunable parameter for this calculation. By adjusting $\Delta E_c$, we can control the "ionization" energy for trapped electrons.

Quantum computing errors that can be repaired fall into two categories: decoherence errors, characterized by a coherence time, $t_f$, and gating errors that accumulate over many operations. Fault-tolerant techniques have been developed for correcting the errors, but these are only effective for error levels up to $10^{-4}$, or one accumulated error per $10^4$ operations[16]. In the coded qubit scheme, a two-qubit operation (like C-NOT) is composed of a sequence of order 10 exchange coupling primitives (like SWAP)[12]. The error level for such primitives should then be about $10^{-5}$, at least until error correction techniques can be optimized for this scheme.

The decoherence time $t_f$ associated with spin-phonon relaxation is large for electron spins in Si. For donor-bound electrons in bulk Si at low temperatures and fields, the spin-lattice relaxation time $T_1$ can be greater than 3000 s[17]. When uniaxial strain is applied, $T_1$ grows by many orders of magnitude[18]. Transverse spin relaxation, $T_2$, is a greater source of concern. On short time scales, there occurs a rapid but very incomplete dephasing, which does not interfere with error correction techniques[19]. Complete dephasing, which occurs over much longer time scales, involves several contributions that can be managed through engineering. We can utilize isotopically enriched $^{28}$Si to reduce hyperfine couplings with $^{29}$Si nuclear spins[20]. Compensation techniques, developed for nuclear magnetic resonance[21], may be able to offset the dipole-dipole interactions between trapped electrons, though this remains controversial[18,22]. Increasing the barrier thickness curbs interactions with electrons in the tunnel-coupled back-gate. The limiting process for $T_2$ in a bulk Si device may therefore be the spin-lattice interaction[19], such that $t_f \cong T_2 \cong T_1$. Confirmation of this point poses an important experimental challenge.

Accurate gating involves two steps: initial characterization of the exchange coupling between pairs of qubits, and precise implementation of the gate operations. We do not pursue the first issue here, although it forms an important and interesting field for future research. As a prototype for gate operations we consider $\sqrt{\text{SWAP}}$, as implemented with a pulse signal, $V_s(t)$. The particular shape of $V_s(t)$ is arbitrary, although the integrated pulse must satisfy the relation[6]

$$\int_{t_s} J(V_s(t))dt = p\hbar/2 \qquad . \tag{1}$$

Here, $t_s$ is the switching time, and the function $J(V)$ was computed in Fig. 2. Fault-tolerant error correction requires that Eq. 1 should be satisfied to an accuracy of $10^{-5}$.

The control pulse $V_s(t)$ can never be implemented perfectly. The results from Fig. 2 allow us to specify performance criteria for control electronics. An exchange pulse of fixed area (see Eq.1) can be shaped low and flat, such that errors in the pulse width are diluted to acceptable levels. For pulse width errors of 100 ps, available from commercial GHz pulse generators[23], error correction requires that the pulse length $t_s > 10$ μs. Faster operation is possible, theoretically, but depends on tighter control of the pulse width. In addition to pulse width errors, we must also avoid gating errors associated with non-adiabatic switching: the exchange coupling should be turned on slowly[24,25]. For a flat-top pulse of width 10 μs, we estimate a minimum pulse edge of 10 ps. The flat-top shape is therefore realistic.

What are the error levels that can be tolerated in the applied gate voltages? The answer to this question is implementation-specific, and can only be determined via modeling. For a flat-top pulse of height $J = p\hbar/2t_s$, fault tolerant error correction requires a pulse height variation of

$$dV < 10^{-5} J / |\partial J / \partial V|. \tag{2}$$

To a good approximation, the function $J(V)$ is exponential, $J \cong J_0 e^{-V/V_0}$, leading to the constraint $dV < 10^{-5} V_0$. Large nonlinearities in $J(V)$ correspond to smaller $V_0$ and tighter voltage control requirements. Fitting the exponential form for $J$ to different regions in Fig. 2 yields $dV < 20-55$ nV, or $dV/V < 5-8 \times 10^{-7}$. For sub-kHz pulses (approaching DC), extremely high voltage accuracy can be achieved. For sub-MHz pulse generators, the desired accuracy levels fall nearly within the specifications of off-the-shelf electronics[26]. For GHz operation, significant improvements will be required before Eq. 2 is satisfied.

Our calculations show that the key challenge for solid-state spin-based quantum computation is to develop devices in which the exchange coupling is insensitive to gate voltage uncertainty. At a simple level, the quantum dot structures should be optimized to increase $V_0$, which sets the scale for gate voltage accuracy requirements. The ultimate goal should be to "digitize" the gating function $J(V)$. A possible solution could employ a bistable quantum dot design in which $J$ is nearly zero over some range of gate voltage, and finite but constant in a different voltage range.

## Methods

The electrostatics calculations are performed using finite-element software. Image potentials arising from the nontrivial gate structure are calculated self-consistently using Green's functions. The quantum mechanical problem is approximated as a single envelope function, due to the absence of strong conduction valley-coupling mechanisms. A basis set of 18 single-electron Hartree-Fock wavefunctions is obtained in real space on an adaptive finite-element mesh. A basis of about 50 two-electron wavefunctions is constructed in the configuration interaction approach. The Hamiltonian matrix is then computed and diagonalized, giving an essentially exact result for the wavefunction.

## Acknowledgements
We have benefited from helpful discussions with C. L. Brace, S. Coppersmith, X. Hu, and C. Tahan. R. Nelson and E. Blevis provided invaluable technical support with the PDE modeling software, FlexPDE©. Our work was supported by the U.S. Army Research Office through the ARDA program, and the National Science Foundation through the MRSEC and QuBIC programs.

## References
1. Shor, P. W. In *Proc. 35th Annu. Symp. Foundations of Computer Science*, (ed. Goldwasser, S.) 124–134 (IEEE Computer Society, Los Alamitos, CA, 1994).
2. Preskill, J. Reliable quantum computers. *Proc. R. Soc. London A* **454**, 385–410 (1998).
3. Kane, B. A silicon-based nuclear spin quantum computer. *Nature* **393**, 133–137 (1998).
4. Mozyrsky, D., Privman, V. & Glasser, M. L. Indirect interaction of solid-state qubits via two-dimensional electron gas. *Phys. Rev. Lett.* **86**, 5112–5115 (2001).
5. Vrijen, R. *et al*. Electron-spin-resonance transistors for quantum computing in silicon-germanium. *Phys. Rev. A* **62**, 012306 (2000).
6. Loss, D. & DiVincenzo, D. P. Quantum computation with quantum dots. *Phys. Rev. A* **57**, 120–126 (1998).
7. Burkard, G., Loss, D. & DiVincenzo, D. P. Coupled quantum dots as quantum gates. *Phys. Rev. B* **59**, 2070–2078 (1999).
8. Levy, J. Quantum-information processing with ferroelectrically coupled quantum dots. *Phys. Rev. A* **64**, 052306 (2001).
9. Kouwenhoven, L. P. *et al*. in *Mesoscopic Electron Transport* (eds. Sohn, L. L., Kouwenhoven, L. P. & Schön, G.) 105–214 (NATO ASI Ser. E Vol. 345, 1997).
10. Mooney, P. M. *et al.* Relaxed $Si_{0.7}Ge_{0.3}$ buffer layers for high mobility devices. *Appl. Phys. Lett.* **67**, 2373–2375 (1995).
11. Bacon, D., Kempe, J., Lidar, D. A. & Whaley, K. B. Universal fault-tolerant quantum computation on decoherence-free subspaces. Phys. Rev. Lett. **85**, 1758-1761 (2000).
12. DiVincenzo, D. P., Bacon, D., Kempe, J., Burkard, G. & Whaley, K. B. Universal quantum computation with the exchange interaction. *Nature* **408**, 339–342 (2000).
13. Hu, X. & Das Sarma, S. Hilbert-space structure of a solid-state quantum computer: Two-electron states of a double-quantum-dot artificial molecule. *Phys. Rev. A* **61**, 062301 (2000).
14. Sham, L. J. & Nakayama, M. Effective-mass approximation in the presence of an interface. *Phys. Rev.* B **20**, 734–747 (1979).
15. Koiller, B., Hu, X. & Das Sarma, S. Exchange in silicon-based quantum computer architecture. *Phys. Rev. Lett*. **88**, 027903 (2002).
16. Nielsen, M. A. & Chuang, I. L. *Quantum Computation and Quantum Information* (Cambridge Univ. Press, Cambridge, 2000).
17. Feher, G. & Gere, E. A. Electron spin resonance experiments on donors in silicon. II. Electron spin relaxation effects. *Phys. Rev*. **114**, 1245–1256 (1959).
18. Tahan, C., Friesen, M. & Joynt, R. Decoherence of electron spin qubits in Si-based quantum computers. Preprint cond-mat/0203319 at (http://xxx.lanl.gov) (2002).
19. Mozyrsky, D., Kogan, S & Berman, G. P. Time scales of phonon induced decoherence of semiconductor spin qubits. Preprint cond-mat/0112135 at (http://xxx.lanl.gov) (2002).

20. Gordon, J. P. & Bowers, K. D. Microwave spin echoes from donor electrons in silicon. *Phys. Rev. Lett.* **1**, 368–370 (1958).
21. Vandersypen, L. M. K. *et al*. Experimental realization of Shor's quantum factoring algorithm using nuclear magnetic resonance. *Nature* **414**, 883–887 (2001).
22. De Sousa, R. & Das Sarma, S. Electron spin coherence in semiconductor quantum computers. Preprint cond-mat/0203101 at (http://xxx.lanl.gov) (2002).
23. Estimates are based on optimal specifications from the Agilent Technologies 8133 and 81100 families of GHz pulse generators (http://www.agilent.com). Errors in pulse height are listed as $dV/V < 10^{-2}$, and pulse widths as $dt < 100$ ps.
24. Burkard, G., Engel, H. A. & Loss, D. Spintronics and quantum dots for quantum computing and quantum communication. *Fortschr. Phys.* **48**, 965–986 (2000).
25. Hu, X. & Das Sarma, S. Gate errors in solid state quantum computer architectures. Preprint cond-mat/0202152 at (http://xxx.lanl.gov) (2002).
26. Specifications for pulse amplitude jitter in PB-4 and PB-5 sub-MHz pulse generators from Berkeley Nucleonics Corporation are listed as $dV/V < 10^{-5}$ (http://www.berkeleynucleonics.com).